# Scanning tunneling microscopy of the 32 K superconductor $(Sr_{1-x}K_x)Fe_2As_2$

M. C. Boyer[1], Kamalesh Chatterjee[1], W. D. Wise[1], G. F. Chen[2], J. L. Luo[2], N. L. Wang[2], E. W. Hudson[1]

[1] Department of Physics, Massachusetts Institute of Technology, Cambridge, MA 02139, USA.
[2] Beijing national Laboratory for Condensed Matter Physics and Institute of Physics, Chinese Academy of Sciences, Beijing 100190, China

**The discovery of high temperature superconductivity in $La[O_{1-x}F_x]FeAs$ at the beginning of this year[2] has generated much excitement and has led to the rapid discovery of similar compounds with as high as 55 K transition temperatures[3]. The high superconducting transition temperatures are seemingly incompatible with the electron-phonon driven pairing of conventional superconductors, resulting in wide speculation as to the mechanism and nature of the superconductivity in these materials. Here we report results of the first scanning tunneling microscopy study of the 32 K superconductor $(Sr_{1-x}K_x)Fe_2As_2$. We find two distinct topographic regions on the sample, one with no apparent atomic corrugation, and another marked by a stripe-like modulation at double the atomic periodicity. In the latter the stripes appear to modulate the local density of states, occasionally revealing a $\Delta = 10$ mV gap with a shape consistent with unconventional (non-s wave) superconductivity.**

Like the cuprate superconductors, the currently known FeAs superconductors are derived from magnetically ordered non-superconducting parent compounds, $LnFeAsO$ (where $Ln$ is one of several of the Lanthanoids) and $AFe_2As_2$ ($A$ = Sr, Ba). In this case the magnetic order is a spin density wave (SDW), which forms below a structural transition at a temperature $T_S \sim 150$ K – 200 K depending on the compound, and is apparent in resistivity, magnetic susceptibility, specific heat, and neutron scattering, among other probes[4-8]. To become superconducting these parent compounds must be doped, typically by depletion or partial substitution of O (e.g. by F) in $LnFeAsO$ and by substitution of $A$ (e.g. by K) in $AFe_2As_2$, leading to superconducting transition temperatures currently as high as 55 K (Ref. 3) and 38 K (Ref. 9-11) respectively. Doping not only adds hole carriers[11,12], but also reduces the signature of the SDW transition, with superconductivity typically arising as the SDW transition vanishes, suggesting competition between the two states.

One possible difference between the cuprates and FeAs materials is the symmetry of the order parameter. While the cuprates are d-wave superconductors[13], the pairing symmetry of the FeAs materials is the subject of intense theoretical[14-26] and experimental[27-35] debate.

Here we weigh in on this debate with the results of scanning tunneling microscopy (STM) of $(Sr_{1-x}K_x)Fe_2As_2$ (Sr-122), doped to a transition temperature $T_C = 32$ K. The samples were grown by a flux technique discussed elsewhere[36]. We cleave the samples in ultra-high vacuum at low temperatures (10 K) and insert them immediately into the instrument and further cool to 5.3 K, at which all data reported here is obtained. It is unclear how cleavage works in Sr-122, but as can be seen in the schematic unit cell of Fig. 1, there are two likely possibilities. Because of the strong bonding between Fe and As those atoms are unlikely to be disturbed. Instead the sample is likely to cleave either between Sr and As planes, yielding two distinct, strongly charge imbalanced planes (as is observed[37] in the cuprate superconductor $YBa_2Cu_3O_{6+x}$), or in the Sr plane, leaving about half of the Sr on each of the two exposed, and roughly charge balanced, surfaces.

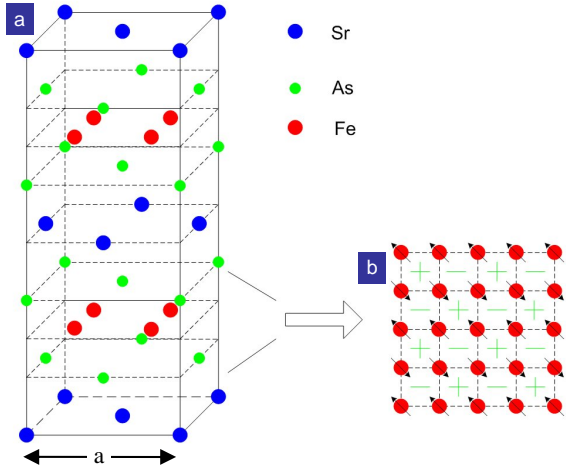

Figure 1. **Schematic unit cell of $(Sr_{1-x}K_x)Fe_2As_2$.** (a) A perspective view of the unit cell – a $ThCr_2Si_2$-type structure – shows Sr/K atoms (blue) aligned with As (green), but effectively rotated relative to the square lattice of the Fe atoms (red). This is clear in the (b) top-down view, which also indicates the orientation of the antiferromagnetic stripe order in the undoped parent compound. Note the effective rotation of the lattice of As atoms, whose positions alternate from above to below a given Fe plane (+ and – sites). Lattice constants: a,b ~ 5.6 Å (so As-As distance is 3.96 Å), c = 12.39 Å (Ref. 1)

Topographic imaging of freshly cleaved surfaces reveals two distinct topographic signatures. On part of the sample we see a disordered surface with no visible lattice or periodic features. These regions are bounded by 13(1) Å step edges, consistent with the c-axis unit cell size, leading to other regions in which we see a strong stripe structure modulating a square atomic lattice (Fig. 2). The atomic spacing along the stripes is 4.0 Å, consistent with the atoms being either Sr/K or As, and the stripe wavelength is twice that, as every other row of atoms is raised relative to its neighbors. The stripes appear to involve a slight dimerization, with atomic rows spaced 3.6 Å from their partner and 4.4 Å from the neighboring row.

In some places the atoms making up the stripe appear to be missing (dark region of Fig. 2b), allowing a view through to the next atomic plane, 2.8(4) Å below. That plane consists of a non-stripe modulated square lattice with atoms spaced and shifted by half a unit cell, but not rotated, from the atoms along the stripes.

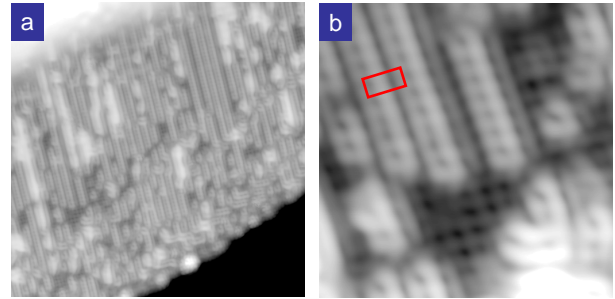

Figure 2. **Topography reveals stripes.** (a) 300 Å constant current topography reveals a stripe ordered region of the sample, bounded by unit cell height step edges in the upper left and lower right leading to non-atomically resolved regions. (b) 30 Å field of view more clearly showing resolution of atoms, which form a dimerized 4.0 Å x 8.0 Å cell (red rectangle). Also visible (dark area, lower center) is the underlying square lattice (4.0 Å square unit cell, offset from the surface lattice by half a unit cell), consistent with the two layers being Sr/K and As respectively. Tunneling parameters: $V_{sample}$ = -100 mV, $I_{set}$ = 200 pA for all data reported here.

These observations lead us to the conclusion that the striped surface consists of a nearly complete Sr/K layer, sitting upon an As layer. The in-plane offset of the lattice below the stripe plane, apparent in the figure, is consistent with this interpretation. Although our measured distance between these two layers – 2.8(4) Å – is larger than their known spacing in the bulk (2.0 Å), this difference is both difficult to assess given the large (nearly 1 Å) amplitude of the stripe modulation and unsurprising given the unbalanced charge expected of a surface Sr/K layer. The stripes themselves are likely to be a surface reconstruction. Although the parent compound is known to have a stripe-like SDW and it is not unreasonable that some remnant of that state would remain in this doped compound, the SDW stripes are oriented along the $(\pi, \pi)$ direction, aligned with the Fe lattice, and hence rotated 45 degrees relative to the stripes we observe.

That the non-atomically resolved layers appear to be about one unit cell from the stripe layer indicates that they are likely Sr/K layers as well, although apparently cleaved differently or perhaps unreconstructed. Without resolving individual atoms in this region it is difficult to say more from the topography.

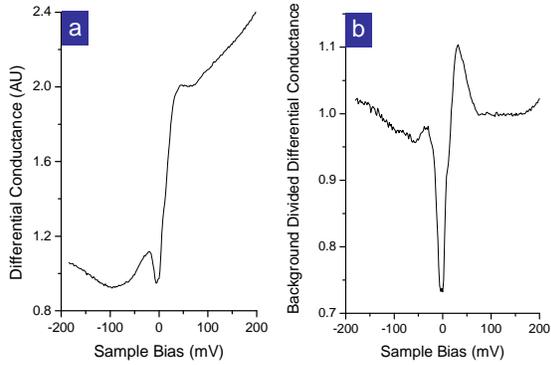

Figure 3. **Spectroscopy on non-atomically resolved regions.** (a) Spectra from non-atomically resolved regions show a strongly asymmetric background. A 32 mV gap can be more clearly observed by dividing out a broadly smoothed background (b).

Having a tentative identification of the surfaces into which we are tunneling, we next proceed to spectroscopy. The differential conductance measured by scanning tunneling spectroscopy is to first order proportional to the local density of states of the sample at the tip location, providing a powerful insight into the spatial and energetic distribution of electronic excitations in the system. We find four distinct spectra associated with well defined spatial features.

First, on non-atomically resolved planes we find a strongly asymmetric background (Fig. 3a) which is often, though not always, gapped. This is reminiscent of spectroscopy on underdoped cuprates, although in those materials the conductance is typically higher at negative sample bias rather than at the positive sample bias seen here. We clarify the gap by dividing out the background (obtained by broadly smoothing the spectrum) and find a 32 mV gap (Fig. 3b). This is comparable to the ~ 40 meV gap observed by angle resolved photoemission (ARPES) in undoped $BaFe_2As_2$ (Ref. 38), and somewhat larger than the 20 meV pseudogap identified in $La(O_{1-x}F_x)FeAs$ by angle integrated photoemission[28,39]. For this reason we tentatively identify this gap as a pseudogap, perhaps associated with the SDW of the parent compound.

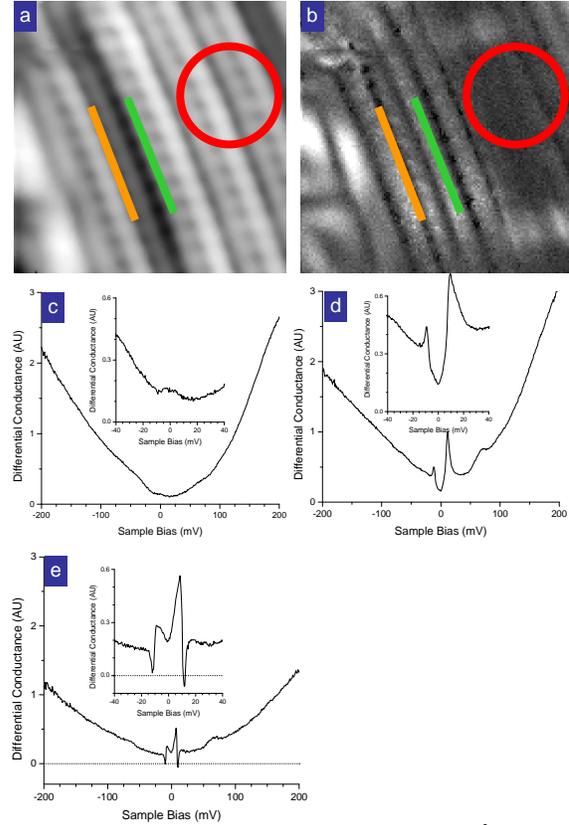

Figure 4. **Stripe modulated spectra**. (a) A 50 Å topographically striped region shows strong modulation of the local density of states, evidenced in (b) a +12 mV conductance map of the same region. Three different classes of spectra are observed, including (c) bowl shaped spectra typical of the region marked with a red circle of (a,b), (d) gapped spectra with sharp coherence peaks near ±10 mV tentatively identified as the superconducting gap (typical of spectra along lines like the green line of (a,b)) and (e) unusual spectra evidencing negative differential conductance peaks at ±11 mV (typical of spectra along lines like the orange line of (a,b)).

The picture is more complex on the striped surface, with the stripes modulating the local density of states. In Fig. 4 we show a 50 Å image of a striped region along with a 12 mV conductance map showing strong periodic modulations of the differential conductance. Overall the spectra from these regions can be segregated into three classes. The first (Fig. 4c) is a broad, nearly featureless bowl shaped spectrum. As these spectra are found centered on the most well ordered stripes (e.g. the region circled in red) and have no strong low energy scale they seem likely to be associated with a insulating-like charge gap.

The other two spectral types appear to broadly share this spectrum as a background, but include additional low energy features. First we find a clearly gapped spectrum with sharp coherence peaks at 10 mV (Fig. 4d). These spectra appear in brighter regions of the conductance map (e.g. the green line of Fig. 4b), corresponding to the lower partner of the striped pairs. The shape of the spectrum is strongly reminiscent of the d-wave gap measured by STM in the cuprates[40]. Although the background is not nearly as asymmetric as that found in the non-atomically resolved layers (Fig. 3a), the positive-bias coherence peak is typically stronger. The gap size is larger than the ~ 4 meV superconducting gap measured in doped $Ln$FeAsO compounds by specific heat[27], point-contact spectroscopy[31], lower critical field measurements[30] and photoemission[28]. We need to await confirmation by similar measurements on Sr-122 to see if this increase in gap size is universally found. If so, then the ratio $2\Delta/k_BT_C \sim 7$ is about a factor of two larger than in conventional superconductors, similar to reported ratios of 8 - 12 in the cuprates, and suggestive of unconventional superconductivity.

In addition to determining its size we can also comment on the pairing symmetry based on the shape of the spectrum. It is clearly not consistent with an isotropic s-wave gap such as is found in conventional superconductors, where no excitations are found below the gap edge. Instead, its v-shape is indicative of a gap with nodes, as found in a p-wave or d-wave superconductor. Although some measurements, most notably microwave penetration depth[35], present evidence for fully gapped superconductivity, most theoretical and experimental studies now lean towards unconventional, either p-wave or d-wave, superconductivity[14-34].

This 10 mV energy scale also appears in the final class of spectra in the striped plane, an unusual spectrum which, after peaking near ±9 mV, shows negative differential conductance peaks near ±11 mV. Although negative differential conductance has been reported in STM of a variety of low dimensional systems[41,42] we are unaware of any such findings in superconducting systems. As can be seen by the locations of the darkest lines in the conductance map of Fig. 4b (e.g. the orange line), these spectra are found on the edges of the brightest stripes, with the negative conductance being strongest roughly halfway between atoms. Further experimental and theoretical investigation is warranted in order to determine the cause of these sharp features and their relation to the presumed superconducting gap of the same energy.

**Acknowledgments**
We thank J.E. Hoffman, P.A. Lee, S. Sachdev, T. Senthil for helpful comments. This research was supported in part by a Cottrell Scholarship awarded by the Research Corporation and by the MRSEC and CAREER programs of the NSF.

**Author contributions**
M.C.B., K.C. and W.D.W. shared equal responsibility for all aspects of this project from instrument construction to data collection and analysis and manuscript preparation. G.F.C, J.L.L and N.L.W. contributed to sample growth. E.W.H advised.

**Author Information**
Correspondence and requests for materials should be addressed to E.W.H.